\documentclass[10pt,journal,cspaper,compsoc]{IEEEtran}
\usepackage{xcolor}
\usepackage{color}
\usepackage{listings} 
\usepackage{courier} 
\usepackage{caption}
\usepackage{amsmath} 
\usepackage{chngpage} 
\usepackage{epsfig}
\usepackage{graphicx}
\usepackage{epstopdf}
\usepackage{graphicx}
\usepackage{epstopdf}

\definecolor{mygray}{HTML}{EDEDED}
\definecolor{mygreen}{rgb}{0,0.6,0}

\lstnewenvironment{code}[1][]%
  {\minipage{\linewidth}
\lstset{
       tabsize=2,
       backgroundcolor=\color{mygray},
       commentstyle=\color{blue},
       basicstyle=\ttfamily\scriptsize ,
       frame=single,
       columns=fullflexible,
       #1
}}
{\endminipage}

\lstdefinelanguage{aspectj}
{
morekeywords={public, void, aspect, pointcut, execution, this, around},
sensitive=false,
morecomment=[l]{//},
morecomment=[s]{/*}{*/},
morestring=[b]",
}

\begin{document}

\title{COARA: Code Offloading on Android with AspectJ}

\author{
	\IEEEauthorblockN{Roy Friedman\IEEEauthorrefmark{1} and Nir Hauser\IEEEauthorrefmark{1} \\}
   	\IEEEauthorblockA{\IEEEauthorrefmark{1} Technion -- Israel Institute of Technology\\
       Department of Computer Science\\
       Haifa, Israel}
}

\maketitle
\begin{abstract}

Smartphones suffer from limited computational capabilities and battery life. A method to mitigate these problems is \emph{code offloading}: executing application code on a remote server. We introduce COARA, a middleware platform for code offloading on Android that uses aspect-oriented programming (AOP) with AspectJ.  AOP allows COARA to intercept code for offloading without a customized compiler or modification of the operating system. COARA requires minimal changes to application source code, and does not require the application developer to be aware of AOP.  Since state transfer to the server is often a bottleneck that hinders performance, COARA uses AOP to intercept the transmission of large objects from the client and replaces them with object proxies. The server can begin execution of the offloaded application code, regardless of whether all required objects been transferred to the server. We run COARA with Android applications from the Google Play store on a Nexus 4 running unmodified Android 4.3 to prove that our platform improves performance and reduces energy consumption. Our approach yields speedups of 24x and 6x over WiFi and 3G respectively.

\end{abstract}

%
%

\section{Introduction}

Mobile computing is quickly becoming the prominent computing and communication platform. Smartphones and tablets have become ubiquitous, while mobile applications are increasing in complexity. However, even with recent improvements in mobile processing power, server processors are substantially more powerful \cite{Flinn:2012ws}.  The primary focus of mobile design is long battery life, minimum size and weight, and heat dissipation.  Computing power is often compromised in favor of these other priorities \cite{Ha_JustInTime:2012wx}.  New mobile devices such as the Pebble watch and Google Glass may have even less computing power than current smartphones \cite{Fahim:2013dd}.

This challenge has been tackled in the past using the client-server model, where the developer is responsible for writing code that issues requests to an API on a server. The downside of this approach is that rather than focusing on application functionality, the developer must invest resources in writing code to support network communication.  If network connectivity is poor or unavailable, the application may be unable to execute.

One approach to addressing these challenges is \emph{code offloading}: using middleware to automatically execute application code on a remote server.  By identifying computation intensive code to be offloaded, we can leverage the disparity between mobile and server computing power to achieve better performance and longer battery life on the mobile device.

In this paper, we introduce COARA, a middleware platform for code offloading on Android that uses aspect-oriented programming (AOP) \cite{AOP_Paper:1997vp} with AspectJ \cite{aspectj:Online}.  AOP allows COARA to intercept code for offloading without a customized compiler or modification of the operating system.  COARA requires minimal changes to application source code.  The application developer does not need to be aware of AspectJ to use COARA.

Execution of a method on a remote sever requires that the application state must be transferred along with the method. State transfer to the server is often a bottleneck that hinders performance, especially on low-bandwidth 3G connections.  To minimize state transfer, COARA uses AOP to intercept the transmission of large objects from the client to the server, replacing the objects with \emph{object proxies}.  Object proxies act as server-side placeholders for objects whose state, by design, has not been transferred to the server. The server can begin execution of the offloaded application code, regardless of whether all required objects have been transferred to the server.

COARA enables two data transmission strategies that use object proxies: \emph{lazy} and \emph{pipelined transmission}. With lazy transmission, COARA assumes that large objects will not be needed and therefore replaces the objects with proxies when transferring state to the server.  If code executing on the server accesses the proxy, COARA will halt execution and grab the object from the mobile device.  With pipelined transmission, COARA also replaces objects with proxies, however operates under the assumption that most of the objects \emph{will} be accessed.  Therefore, immediately after offloading the method, COARA transmits the large objects to the server in a pipelined manner.  Pipelined transmission allows the server to get a ``head start'' and execute code while the rest of the heap is still being transferred.  COARA also supports an \emph{eager transmission} strategy that transfers the entire reachable heap to the server without the use of object proxies. We discuss these strategies in greater detail in Section \ref{section.object_proxies} and show their effectiveness in improving performance in our evaluation.

We evaluated COARA using a Nexus 4 running unmodified Android 4.3 on applications from the Google Play store as well as applications we wrote that wrap open source Java libraries. Our approach achieves speedups of 24x and 6x over WiFi and 3G respectively. While this paper focuses on performance improvement, COARA was also able to achieve reductions of energy consumption of 11x over WiFi and 3x over 3G.

The rest of this paper is organized as follows.
We discuss related work in Section~\ref{sec:related} and present some background terminology and technology in Section~\ref{sec:background}.
The design goals and architecture of COARA are explained in Section~\ref{sec:arch}.
We evaluate the usage and performance of COARA in Section~\ref{sec:evaluation}.
We discuss limitations of the current system and future work in Section~\ref{sec:limits} and conclude in Section~\ref{sec:conclusion}.

\section{Related Work}
\label{sec:related}
FarGo \cite{Holder:1999wj} is a system that enables distributed applications to maintain a separation between application logic and the layout of components on the underlying physical system. FarGo allows components known as \emph{complets} to be relocated based on events at runtime.  FarGo extends Java and uses Java RMI to communicate between JVMs.  While FarGo is able to decouple application logic from layout logic, the system is tightly coupled with the application itself.  FarGo requires that existing applications be rewritten to adhere to the FarGo programming model.

Early attempts at offloading like Spectra \cite{Flinn:2001wh} and Chroma \cite{Balan:2003uz} focused on partitioning the application into modules and calculating the optimal offloading strategy.  These partitioning schemes require significant modification of the original application.

The MAUI \cite{Cuervo:2010:MMS:1814433.1814441} architecture provides the capability of fine-grained code offload with minimal application changes.  It shows that minimizing the size of application state transfer vastly improves performance by allowing more methods to be offloaded.  MAUI minimizes state transfer by only sending the difference in application state on subsequent method call invocations. However, MAUI serializes the application state using XML which is much slower than binary serialization and results in data that is much larger \cite{Hericko:2003tk}.  MAUI uses a custom compiler to insert offloading logic into an application, which generates two separate code bases --- one for the mobile device and one for the server.  Custom compilation may complicate debugging for the application developer.

CloneCloud's \cite{Chun:2011ty} approach is to migrate the Virtual Machine (VM) on which the application is running.  It migrates the VM from an Android phone to a server. Unlike other approaches, CloneCloud does not require intervention on the part of the developer for offloading since it happens at the operating system level.  It also has the ability to offload native method calls.  The downside of CloneCloud's approach to offloading is the overhead required to migrate the entire VM from the mobile device to the server.  In addition, the server must already be running a hardware simulator with the same configuration as the mobile device, further complicating matters.

ThinkAir \cite{Kosta:2012vx}, like MAUI, provides method level code offloading but focuses more on scalability issues and parallel execution of offloaded tasks.  It uses a custom compiler to insert offloading logic.  ThinkAir creates a VM on the cloud and can allocate multiple VMs to handle a single application.  However, the authors do not discuss how they merge the application states from multiple VMs back to the original application.

The Cuckoo \cite{Kemp:2012vn} framework takes advantage of the existing activity/service model in Android.  The model makes a separation between the computation intensive parts of the application and the rest of the application.  Cuckoo offloads the well-defined computation intensive portions of the applications.  It requires the developer to write offloadable methods twice --- once for local computations and once for remote computations.  This allows flexibility but may lead to unnecessary code duplication.

COCA \cite{Chen:2012cs} uses AspectJ to offload Android applications to the cloud. COCA finds that the overhead incurred by AspectJ is minimal.  However, COCA is only capable of offloading pure functions which only access the function parameters and are therefore oblivious to the rest of the heap.  Hence, COCA does not tackle the hard problem of state transfer which MAUI demonstrates is one of the key challenges of offloading. Requiring real world applications to only offload pure functions would severely limit the number of functions that could be offloaded.  On the hand, COARA supports state transfer and provides techniques to minimize the resulting overhead.

Calling the Cloud \cite{Giurgiu:2009vf} is a middleware platform that can automatically distribute different layers of an application between the phone and the server, and optimize a variety of objective functions.  However, it requires the application to be partitioned into several inter-connected software modules using the R-OSGi module management tool.  The authors expect that application developers could take up to a month to adapt their applications to be compatible with their solution.

COMET \cite{Gordon:2012vu} is a runtime system that performs offloading with Distributed Shared Memory (DSM).  DSM provides a virtual shared memory space that is accessible by threads on different machines without any work on the part of the developer.  The advantage of this approach is that fine-grained parallel algorithms can be offloaded to multiple machines, resulting in improved performance.  The downside of the DSM approach is that it can lead to large unnecessary state transfers.  For example, COMET requires 4.5MB of state transfer to execute a chess game.  For this reason, COMET is largely unsuccessful in improving performance over low-bandwidth 3G networks.  Recent attempts at distributed memory, such as IBM's X10 programming language \cite{Charles:2005ud}, have forced the developer to be conscious of any remote calls in order to avoid this problem.
 COARA does not currently support offloading parallel algorithms.

Much of the work described in this section focuses on algorithms that decide when to offload and when to execute on the mobile device.  In this paper we focus on \emph{how} to offload code, and less on \emph{when} to offload.  However, COARA does provide a simple decision engine that decides when methods should be offloaded. Naturally, it can be extended with more sophisticated mechanisms.

\section{Background}
\label{sec:background}

\subsection{Android}
Android is a Linux-based operating system designed primarily for touchscreen mobile devices.  Android applications are mostly developed in the Java language using the Android Software Development Kit (SDK) and compiled to Java bytecode. They are then converted from Java Virtual Machine-compatible class files to Dalvik-compatible DEX (Dalvik Executable) files that run on the Android Dalvik Virtual Machine (DVM). During the Android application build process those DEX files are bundled into an APK (Android Application Package) file.  While not all Java bytecode can be translated to DEX, many Java libraries are compatible with Android.

\subsection{Aspect-Oriented Programming (AOP) and AspectJ}
AOP is a programming paradigm that promotes modularity by allowing the separation of \emph{cross-cutting concerns} \cite{AOP_Paper:1997vp}. Concerns are defined as wide-reaching functional or nonfunctional requirements, such as logging, security, or offloading which can be found across modules. Rather than scattering the code to address these concerns throughout the application, we handle them in special classes called \emph{aspects}.  Aspects can alter behavior in the base code by defining a dynamic set of locations in the base code, using a \emph{pointcut}, along with the functionality, or \emph{advice}, that should be added. Our primary cross-cutting concern is offloading the execution of certain methods.

AspectJ \cite{aspectj:Online} is a widely used aspect-oriented extension for the Java programming language. It is open source and is integrated into Eclipse. AspectJ is a natural fit for our architecture.  COARA uses AspectJ to implement an \\\mbox{\emph{OffloadingAspect}} that handles offloading.  COARA identifies which method invocations it wants to intercept with the appropriate pointcut. The use of AspectJ allows COARA's intervention in the application to be expressed by an extension of the Java language in a clear and elegant way.

An additional feature of AspectJ is \emph{inter-type declaration}, which enables aspects to add fields to classes and interfaces.  COARA uses inter-type declaration to implement the \emph{object proxy} transmission optimization by adding a unique global identifier field to objects represented by proxies.

 It is important to note that an application developer does not need to be familiar with or even aware of AOP to use COARA.  However, a developer that would like to modify the COARA source code would find that COARA's business logic is clearly expressed through AOP.

\subsection{Java Remote Method Invocation (RMI)}
Java RMI is a Java API that performs the object-oriented equivalent of remote procedure calls (RPC).  RMI provides the ability to make calls from one JVM to another.  Upon a remote method invocation, all relevant Java objects are automatically serialized and sent to the remote JVM.  RMI allows COARA to send the state of the application from the Android device to the server using native Java binary serialization.
Since Dalvik does not support Java RMI, COARA uses an open source lightweight implementation of RMI called \emph{lipermi} \cite{lipermi:Online}.
\subsection{Java Annotations}
A Java Annotation is metadata that can be added to Java source code.  COARA allows the developer to specify which methods should be offloaded with the use of annotations.  The developer simply tags a method with the annotation \emph{@RemotableMethod} and a decision engine decides whether or not to offload the method at runtime.  Similarly, a developer can tag a class with the \emph{@EnableProxy} annotation to enable \emph{object proxies}.

\section{Design Goals and Architecture}
\label{sec:arch}

The primary goal of COARA is to improve performance by offloading intensive computations while maximizing usability and ease of use for the application developer.  To this end we enumerate the following objectives:

\begin{itemize}
\advance\leftskip-.5cm
  \item \textbf{Simplicity and extensibility of the framework}\\COARA avoids custom compilation and low level operating system changes.  Our offloading strategy is clearly expressed through AOP.
  \item \textbf{Minimal changes to existing code} --- We do not require developers to make significant changes to existing applications, nor do we force new applications to adhere to strict requirements or learn new a programming paradigm.  Incorporating COARA into existing projects should be quick and easy.
  \item \textbf{Minimize state transfer} --- The overhead of state transfer is a significant impediment to the success of offloading, especially over low-bandwidth 3G networks.  COARA offers techniques to minimize state transfer and improve performance.
  \item \textbf{Unmodified Android source code} --- The Android operating system is open source and therefore it is possible to modify the base code. We do not modify the base code to enable application developers to use COARA immediately on any Android device running a recent version of Android.
  \item \textbf{Alternative code execution on server} --- In most cases the server executes the same code as the mobile device.  However, in certain cases we want to increase the \emph{fidelity} of the application by running a more powerful algorithm on a server than we would on the mobile device.  COARA provides the application developer the option to easily specify alternative code.
  \item \textbf{Code privacy} --- Some developers are not comfortable with sharing all of their class files with a remote server.  COARA allows the developer to easily prevent unnecessary class files from being transfered to the server.
\end{itemize}

\begin{figure}
\centering
\includegraphics[width=\columnwidth]{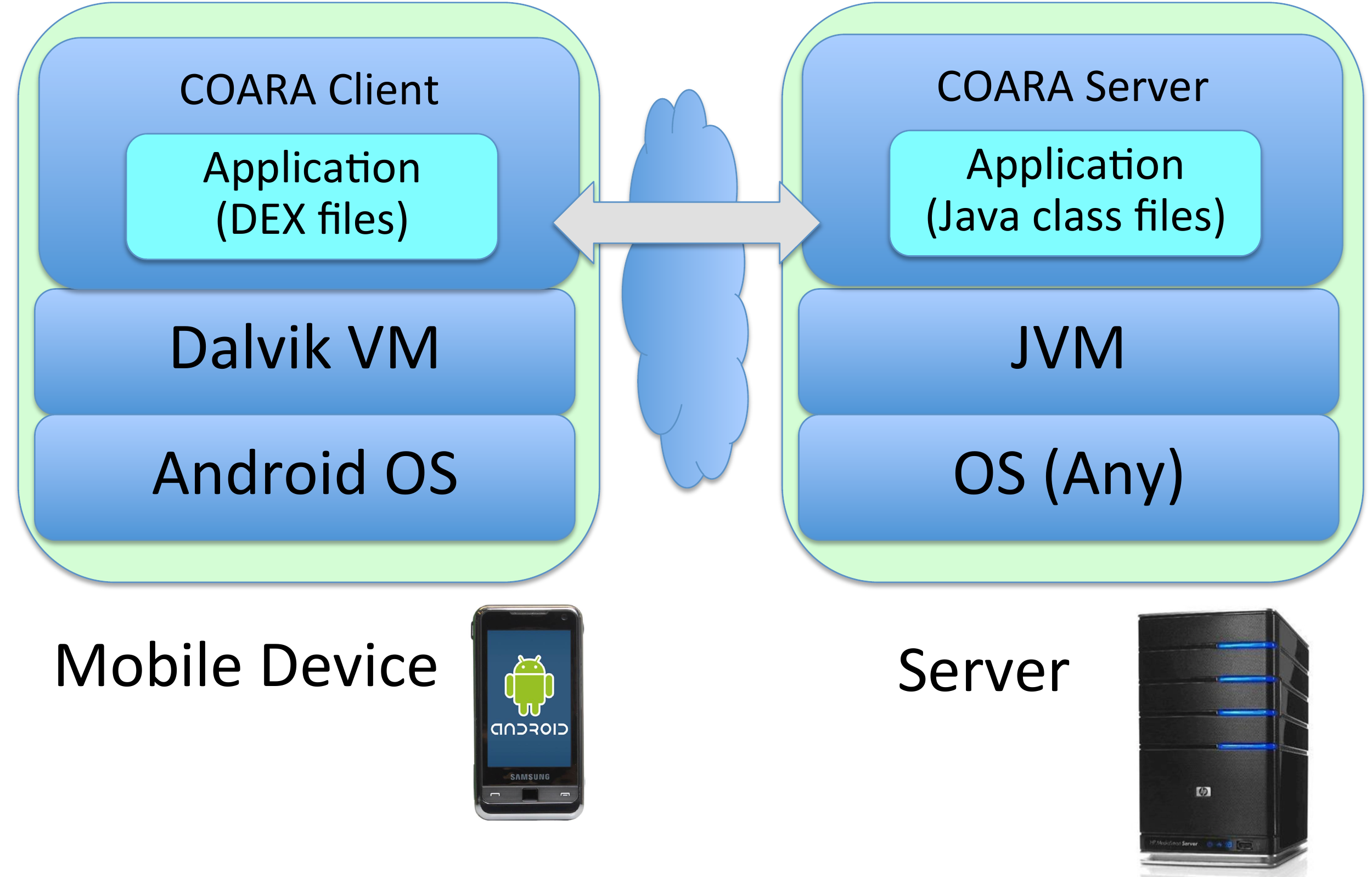}
\caption{\small\textsl{COARA architecture}}
\label{fig:arch}
\end{figure}

\subsection{Code transfer and registration \label{section.code_transfer}}

The COARA architecture is illustrated in Figure \ref{fig:arch}. The COARA server runs within a Java Virtual Machine (JVM).  When a COARA-enabled Android application starts up, the COARA client connects to a fresh instance of a JVM running the COARA Server.  Since Java is a portable language, it does not matter what operating system or architecture the JVM is running on, or whether that operating system is running within a virtual environment on top of a hypervisor.

COARA ensures that the server executes the same application code as the mobile device.  During the Android application build process, a JAR file containing the Java bytecode of the application is bundled within the APK (Android Application Package) file.
When the application begins execution on the mobile device, the COARA client sends an MD5 hash of the contents of the JAR file to the server.  If the COARA server sees the JAR has been previously registered, it loads the JAR.  Otherwise, the server requests the JAR file from the client.  When the server receives the JAR from the client, it loads the JAR  and caches the JAR locally for future sessions.  While this approach adds some overhead at startup, it ensures that the code on the client and server are identical.  To reduce this overhead, it would be possible to introduce a code repository similar to the one found in \cite{Cuervo:2010:MMS:1814433.1814441}.

\subsection{Method identification \label{section.method_idenification}}


In order to identify candidate methods for offloading, \\COARA takes a similar approach to \cite{Cuervo:2010:MMS:1814433.1814441} and \cite{Kosta:2012vx}, and requires the application developer to annotate the offloadable methods with the annotation \emph{@RemotableMethod}.

\begin{figure}
\begin{code}[language=AspectJ]
public aspect OffloadingAspect {

  // Define the annotated methods to intercept (pointcut)
  pointcut executeRemotable(Object obj) :
    execution (@RemotableMethod * *.*(..)) && this(obj);

  // Define execution after interception (advice)
  Object around(Object obj) : executeRemotable(obj) {
    ...
    // Perform offload
    Response response = offloader.executeMethod(
      appState, obj, method, params);
    ...
  }
  ...
}
\end{code}
\caption{\small\textsl{OffloadingAspect.aj}}
\label{fig:OffloadingAspect}
\end{figure}

\begin{figure}
\begin{code}[language=Java]
public class OffloadingServer implements Offloader {
  public Response executeMethod(AppState state, Object obj,
      MethodWrapper method, List<Object> params)
      throws IOException {
    ...
    // Invoke the offloaded method using Java Reflection
    Object returnObject =
      method.getMethod().invoke(obj, params.toArray());
    ...
}
\end{code}
\caption{\small\textsl{OffloadingServer.java}}
\label{fig:OffloadingServer}
\end{figure}


As an example, we have enabled COARA in the open source Google Play application \emph{Pocket Chess For Android} \cite{pocketChess:Online}. The only changes we made to the source code were:
\begin{raggedright}
\begin{enumerate}
  \item Modified \emph{Position} and \emph{SimpleEngine}
  to implement \emph{java.io.Serializable}
  \item Added the \emph{@RemotableMethod} annotation to \emph{SimpleEngine.go()}
    \begin{code}[language=Java]
@RemotableMethod
  public String go() { ... }
       \end{code}
\end{enumerate}
\end{raggedright}

Once one or more methods are annotated, AspectJ will automatically intercept execution and consider it for offload. However, the application developer does not need to be aware of how AspectJ works;  the only intervention required by the application developer is the annotation.

\subsection{Method offload and state transfer \label{section.serialization}}

In order to execute a method remotely, it is not necessary to transfer the entire heap.  The application state which COARA must transfer to the server includes:
\begin{itemize}
 \item static objects in the heap
 \item the method parameters
 \item the object on which the method is invoked, and any objects in the heap that are reachable by that object.  We will refer to this collection of objects as the \emph{reachable heap}.

 \end{itemize}
 Rather than always transferring all static objects, COARA allows the developer to list the objects that are needed specifically in a configuration file.

When a method annotated with \emph{@RemotableMethod} is invoked on the client, the invocation is intercepted by the \\\emph{OffloadingAspect} (Figure \ref{fig:OffloadingAspect}) which invokes the RMI method \\\emph{Offloader.executeMethod(..)} on the COARA server.   When the COARA server receives an offloading request via RMI, \emph{OffloadingServer} (Figure \ref{fig:OffloadingServer}) uses Java reflection to dynamically invoke the method.  In order to allow the offload of any method, the method metadata is passed as a parameter to \emph{Offloader.executeMethod(..)}.  When method execution concludes, the COARA server sends the modified state back to the client.  The COARA client then uses Java reflection to update the transfered objects.

We considered other serialization schemes in addition to native Java Serialization.  We looked at the Kryo \cite{kryo:Online}, XStream \cite{xstream:Online}, and Gson \cite{gson:Online} serialization libraries.  However, we found that those libraries are better at handling pre-defined types.  Since COARA needs to handle an object of any class, we found they did not meet our needs.

\subsection{Object Proxies \label{section.object_proxies}}

\begin{figure}
\includegraphics[width=\columnwidth]{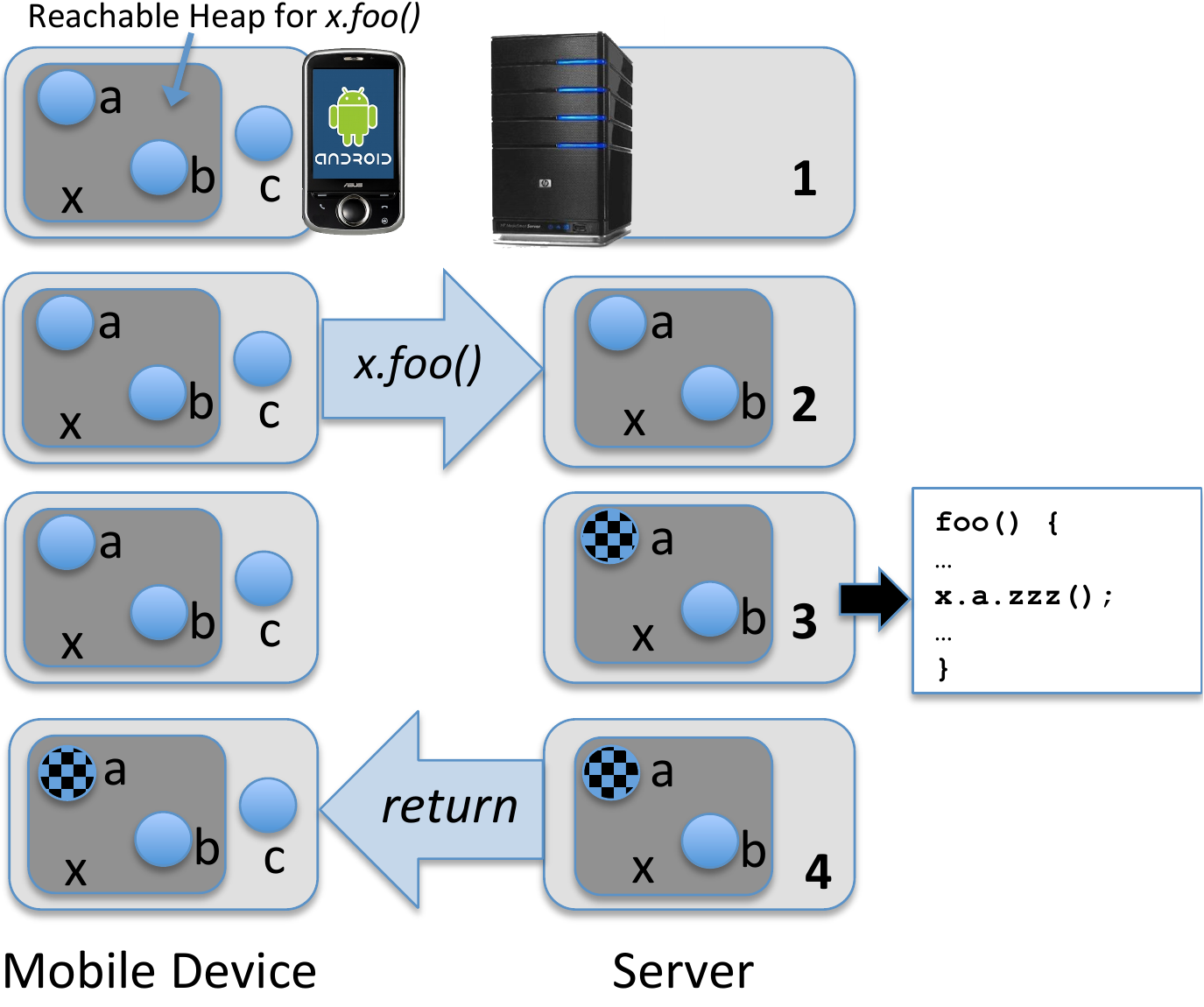}
\caption{\small\textsl{\textbf{eager transmission ---} \\
 1) The object x contains objects a and b.\\
 2) x.foo() is invoked on the client.  x is sent to the server.\\
 3) x.a is modified on the server.\\
 4) x.foo() returns and the state in the client is updated.}}
\label{fig:eager}
\end{figure}

\begin{figure}
\vspace*{-0.6cm}
\includegraphics[width=\columnwidth]{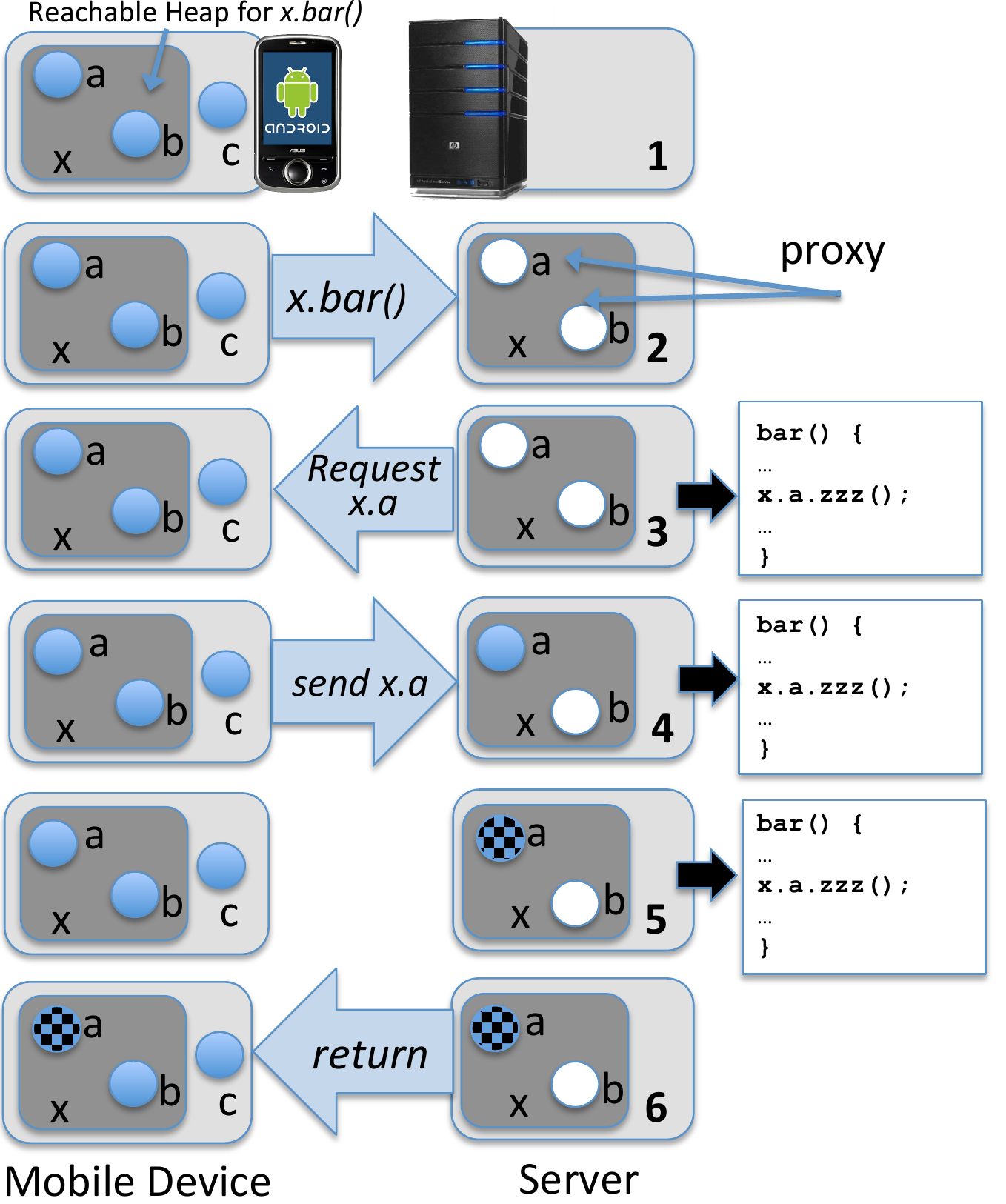}
\captionsetup{width=9cm}
\caption{\small\textsl{\textbf{lazy transmission ---} \\
 1) The object x contains objects x.a and x.b.  \\
 2) x.bar() is invoked on the client.  Proxies are sent for x.a and x.b\\
 3) COARA halts execution on server and requests x.a from client.\\
 4) The client sends x.a to the server.\\
 5) Execution resumes on the server.  x.a is modified.\\
 6) x.bar() returns and the state in the client is updated.}}
\label{fig:lazy}
\end{figure}

For large state transfers, the major components of the state may be large objects that represent media such as images or sounds.  The time required to serialize and transfer the large objects may decrease the speedup gains from offloading, especially over 3G.  COARA mitigates this problem with \emph{object proxies}.

The proxy pattern is a software design pattern.  A proxy is an object that acts as a surrogate or placeholder for another object \cite{DesignPatterns:1995wx}. Proxies are used to defer the cost of transferring an object until the object is needed.  The use of proxies enables the server to begin execution of the offloaded method, regardless of whether all the objects in the \emph{reachable heap} have been transferred.

While the use of object proxies relies heavily on AspectJ, the application developer does not need to be aware of AspectJ to benefit from object proxies.
The application developer enables object proxies on a particular class by annotating the class with \emph{@EnableProxy}. Using AspectJ's inter-type field declaration, COARA adds three fields to the annotated class:

\begin{itemize}
  \item \emph{guid} (Global Unique Identifier) --- used by the server to request an object from the client
  \item \emph{emptyContainer} (boolean) --- \emph{true} if the real object is currently \textbf{not} available in the server's local memory
  \item \emph{isInCache} (boolean) --- true if the real object is currently available in the server's object cache (described in Section \ref{section.cache}).
\end{itemize}

When COARA decides to offload a method to the server, it must first serialize the reachable heap on the client (as described in Section \ref{section.serialization}).  If during serialization an object whose class is annotated with \emph{@EnableProxy} is present in the reachable heap, COARA may replace the object with a proxy --- an empty object instantiated via the default constructor.   The serialized heap containing the proxy is transferred to the server.  When the server deserializes the heap, it will contain a proxy rather than the actual object.
If code executing on the server accesses an object proxy, AspectJ will halt execution until COARA makes the object available.  Once the real object is available on the server, COARA uses reflection to copy the real object's fields into the proxy and normal execution resumes.

We consider three separate strategies when dealing with state transfer from the client to the server: \emph{eager transmission}, \emph{lazy transmission} and \emph{pipelined transmission}. Each strategy is described along with a figure that illustrates a \emph{different} use case.

\begin{itemize}
\advance\leftskip-.5cm
  \item \textbf{\emph{eager transmission}}--- The entire reachable heap is transferred to the server with the offloaded method.  Proxies are not used.  See Figure \ref{fig:eager}.
  \item \textbf{\emph{lazy transmission}}--- Objects whose class is annotated with \emph{@EnableProxy} are replaced with proxies during serialization on the client.  The reachable heap (containing proxies) is transferred to the server with the offloaded method.  If code executing on the server accesses a proxy, AspectJ halts execution and the COARA client issues a synchronous request for the object to the client.  When the request completes, the object is loaded and execution resumes.  See Figure \ref{fig:lazy}.
  \item \textbf{\emph{pipelined transmission}}---  Objects whose class is annotated with \emph{@EnableProxy} are replaced with proxies during serialization on the client.  The reachable heap (containing proxies) is transferred to the server with the offloaded method. As soon as the method is offloaded, the client immediately initiates asynchronous transfers of the real objects to the server, one at a time.  This allows code executing on the server to get a ``head start'' on computation without requiring all of the reachable heap to be present on the server.  We call this strategy ``pipelined'' because the server is simultaneously computing while downloading object data for the next computation.  In addition the client is simultaneously uploading object data while serializing objects for future upload. See Figure \ref{fig:pipelined}.
\end{itemize}

\emph{Lazy transmission} is useful when there are many large objects in the reachable heap, but some may not be accessed by code executing on the server. \emph{Pipelined transmission} is useful when most of the large objects will be accessed by code executing on the server and the application developer has a general idea of the order in which they will be accessed.

COARA uses lazy and pipelined transmission only when transferring state from the client to the server.  If COARA were to use these strategies when transferring state from the server to the client, a server failure would leave the client unable to continue execution because the client would be dependent on the server for a portion of the state.

In Section \ref{sec:evaluation} we show that offloading with object proxies yields speedups of 24.7x for WiFi and 6.1x for 3G and can decrease state transfer by over 90\%.

\section{Decision Engine \label{section.serialization}}

COARA offers a simple decision engine that decides whether to offload a specific method, and if so, which transmission strategy to use.  The decision engine interfaces with a network profiler that periodically measures the bandwidth and latency of the network.  The decision engine can also be overridden programmatically by the application developer.  COARA's architecture is designed to allow a developer to extend COARA and implement an alternative decision engine.  In the future we plan to develop an advanced decision engine.

\subsection{Object Cache \label{section.cache}}

In order to minimize state transfer, COARA provides an object-level cache.  When COARA transfers an object whose class is annotated with \emph{@EnableProxy}, COARA caches a serialized version of the object at the client as well as the server.  The next time COARA serializes the object for transfer, COARA compares the object with the previously stored copy.  If the object has not been modified, COARA replaces the object with a proxy and sets the \emph{isInCache} field to \emph{true}.  When the object is deserialized on the server, COARA will notice that the \emph{isInCache} field is \emph{true} and will retrieve the object from the server's object cache and copy it into the proxy using reflection.

In our evaluation we find that the object cache achieves speedups of 31.9x over WiFi and 23.3x over 3G and can decrease state transfer by 99\% in the best case scenario of 100\% cache hit.

\subsection{Alternative Code execution\label{section.alternative_code_execution}}

Application fidelity can be defined as ``the degree to which the results produced by an application match the highest quality results that would be produced given ample computational resources'' \cite{Flinn:2012ws}. COARA enables application developers to designate alternative method implementations when executed on a remote server in order to increase fidelity.

For example, COARA can run a simple algorithm if the method is executing locally and a computationally expensive algorithm that provides higher quality results if it is executing on a remote server.  The application developer enables this feature for a particular method by defining the optional elements \emph{alternativeClassName} and {\emph{alternativeMethodName} in the \emph{@RemotableMethod} annotation as demonstrated in Figure \ref{fig:alt}.

\begin{figure}
\includegraphics[width=\columnwidth]{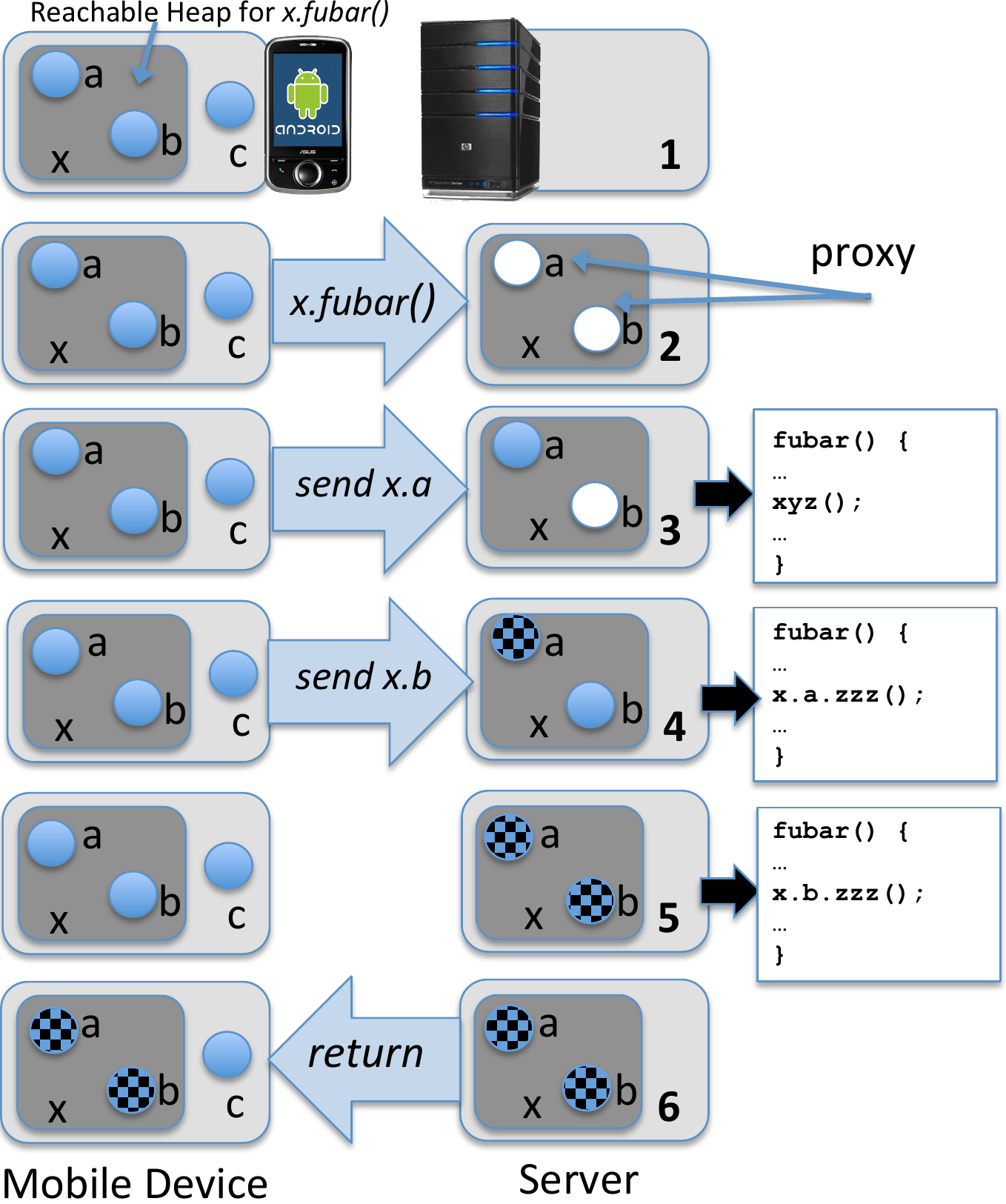}
\captionsetup{width=9cm}
\caption{\small\textsl{\textbf{pipelined transmission ---} \\
 1) The object x contains objects x.a and x.b.\\
 2) x.fubar() is invoked.  Proxies are sent for x.a and x.b.\\
 3) The client immediately sends x.a to the server.\\
 4) The client sends x.b to the server. x.a is modified.\\
 5) x.b is modified.\\
 6) x.fubar() returns and the state in the client is updated.}}
\label{fig:pipelined}
\end{figure}

\begin{figure}
\begin{code}[language=Java]
class MyClass{
  @RemotableMethod (
    alternativeClassName = "com.myApp.MyClass",
    alternativeMethodName = "executeAdvancedAlgorithm"
  )
  public void executeAlgorithm() // Runs on client
    { ... }

  public void executeAdvancedAlgorithm() // Runs on server
    { ... }
}
\end{code}
\caption{\small\textsl{Alternative Code execution code}}
\label{fig:alt}
\end{figure}

\subsection{COARA Build process \label{section.build}}

We designed the COARA client to be easy to integrate into existing Android applications.  The application developer has to follow a series of simple steps to enable COARA in Eclipse:

\begin{enumerate}[noitemsep]
  \item Install the AspectJ Eclipse plugin and enable AspectJ in the application.
  \item Copy the AspectJ runtime JAR (\emph{aspectjrt.jar}) to the \emph{libs} folder.
  \item Add the COARA JAR file to the project's \emph{aspectPath}.
  \item Fill in a COARA XML configuration file (see Table \ref{table:config}).
  \item Add the \emph{android.permission.INTERNET} permission to the Android Manifest.
  \item Add the provided COARA Ant file to the list of Builders for the project (Figure \ref{fig:build.xml}).
\end{enumerate}

The Ant file allows exclusion of sensitive class files that the developer does not want to be offloaded via the \emph{exclude} field.

After these steps are followed, the application developer enables offloading by designating at least one method with the \emph{@RemotableMethod} annotation.

\begin{table}
\centering
\caption{config.xml parameters\label{table:config}}
\begin{tabular}{|r|c|p{4.5cm}|} \hline
\textbf{Parameter}&\textbf{Type}&\textbf{Comment}\\ \hline
server\_ip&string&offloading server IP address\\ \hline
server\_port&int&offloading server port\\ \hline
static\_classes&array&list of static classes to offload \\ \hline
cache\_enabled&boolean&object cache (see Section \ref{section.cache}) \\ \hline
proxy\_enabled&boolean&proxy server enabled\\ \hline
proxy\_ip&string&proxy server IP address\\ \hline
proxy\_port&int&proxy server port\\
\hline\end{tabular}
\end{table}

\begin{figure}
\caption{\small\textsl{build.xml Ant file}}
\begin{code}
<project name="app" default="build-jar" basedir="." >
  <target name="build-jar" >
    <mkdir dir="res/raw"/>
    <jar jarfile="res/raw/classes.jar"
      basedir="bin/classes"
      excludes="**/SecretClass.class"/>
  </target>
</project>
\end{code}
\label{fig:build.xml}
\end{figure}

\subsection{Running Android on the server}

In Section \ref{section.code_transfer}, we indicated that the COARA server runs within a JVM.  However, the JVM is unable to execute Android-specific classes which prevents some methods from being candidates for offloading.  Therefore we have also successfully offloaded code to a COARA server running within an instance of Android-x86 \cite{androidx86:Online}.  We have written a COARA Server Android application which wraps the COARA Runtime and runs within the Dalvik VM as demonstrated in Figure \ref{fig:arch_android}.

\begin{figure}
\centering
\includegraphics[width=\columnwidth]{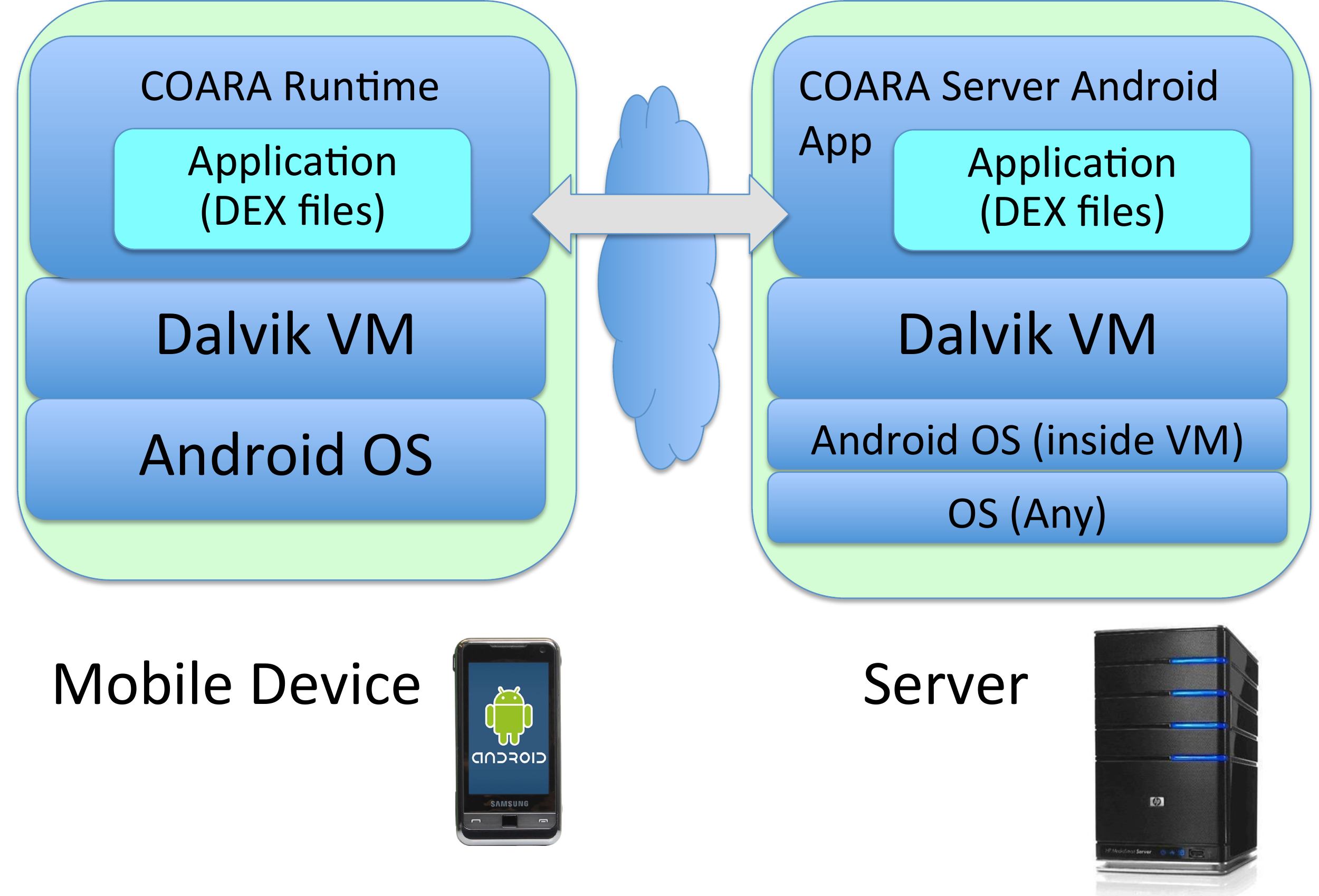}
\caption{\small\textsl{COARA architecture with Android running on server}}
\label{fig:arch_android}
\end{figure}

Unlike \cite{Chen_Androidx86:2010bx} we did not succeed in leveraging the full potential of the server hardware with Android-x86 and therefore did not see significant performance improvement. As our team's ability to extract better performance from Android-x86 improves, we will reconsider running the COARA server exclusively on Android-x86.

\subsection{Error handling}
COARA employs a simple timeout mechanism to handle server failures.  If COARA detects a timeout, it simply runs the method on the local device.  COARA has a Network Profiler that will attempt to reconnect with the server after a failure.  Since any partial computations that may have happened on the server are not visible to the mobile device, there is no risk of corrupting data.  The only risk is if the server executed a transaction on some other network resource which could be duplicated if it were executed again by the mobile device.  Where such a risk exists, the developer should not allow COARA to offload the method.

\subsection{Security}
At the moment, COARA does not provide any security guarantees, therefore the server should be trusted by the client.  In addition all data is transmitted without any encryption.  However, COARA enables the developer to restrict which classes are offloaded to the server.  Further, when the Android device is rooted, the COARA client can communicate with the server through an SSH tunnel.
Adding direct support for encrypted communication using standard means such as SSL is one of our top priorities.

\section{Evaluation}
\label{sec:evaluation}

\subsection{Methodology}
We evaluated the COARA client using an LG Nexus 4 with a \mbox{quad-core} 1.5 GHz Krait CPU running unmodified Android 4.3 JellyBean.  For the offloading server, we used a Dell PowerEdge R210 II with an Intel Xeon E3-1220 V2 3.10GHz single-core CPU running Ubuntu 12.04.3 LTS with JVM 1.6.  We demonstrate that COARA can achieve significant performance gains and energy savings using a commodity server.

For each evaluation workload, COARA used either 802.11g WiFi in the Taub Computer Science building at the Technion Israel Institute of Technology,  or the 3G network of the HotNet cellular provider in Israel.   Mobile devices do not need to be rooted to support COARA.  However, we rooted our Nexus 4 in order to use \mbox{SSH Tunnel} \cite{SSHTunnel:Online} to access a server behind a firewall when testing with 3G.

To evaluate energy consumption, we measured the voltage and current approximately every 10 milliseconds programmatically.  On the Nexus 4, the Android OS exposes these values through the filesystem in files called \emph{current\_now} and \emph{voltage\_now} in the \emph{/sys/class/power\_supply/battery} folder.

We have published the source code for COARA, the example applications, and the automated tests.  The code be found at \emph{https://code.google.com/p/coara/}.

\subsection{Case Studies}

We evaluate the performance and energy consumption of COARA on four Android applications.  Each application was analyzed separately as a case study to more effectively convey the benefits and shortcomings of offloading.

\subsubsection{Pocket Chess}
We evaluated the performance of Pocket Chess, an open source Android application available on the Google Play store \cite{pocketChess:Online}.  In order to be able to run the same game multiple times for this evaluation, we modified the source code of the chess game to remove any randomness in the chess engine.  We then simulated the human moves using Android's \emph{uiautomator} testing framework \cite{uiautomator:Online}.  This resulted in a repeatable 30 move chess game.

While COARA succeeded in improving computation time to determine the next move, the execution time for the User Interface (UI) tasks remained constant as expected.  In Figure \ref{fig:chessPerf} we break down the computation time by UI and Computation. COARA achieved overall game speedups of 3.9x over WiFi and 3.2x over 3G.  If we focus solely on the computation portion, COARA achieved speedups of 9.7x over WiFi and 6.0x over 3G.

When evaluating energy consumption, COARA achieved improvements of 6.4x over WiFi and 3.1x over 3G. For lack of space, an energy consumption graph for Pocket Chess was not included in this version of the paper.

Pocket Chess represents an ideal candidate for COARA due to the combination of high computation and small state transfer.  From a qualitative perspective, without COARA there is noticeable lag whenever the opponent is deciding the next move.  With COARA enabled, it appears to the end user as though the decision is instantaneous.

\begin{figure}
\centering
\includegraphics[width=\columnwidth]{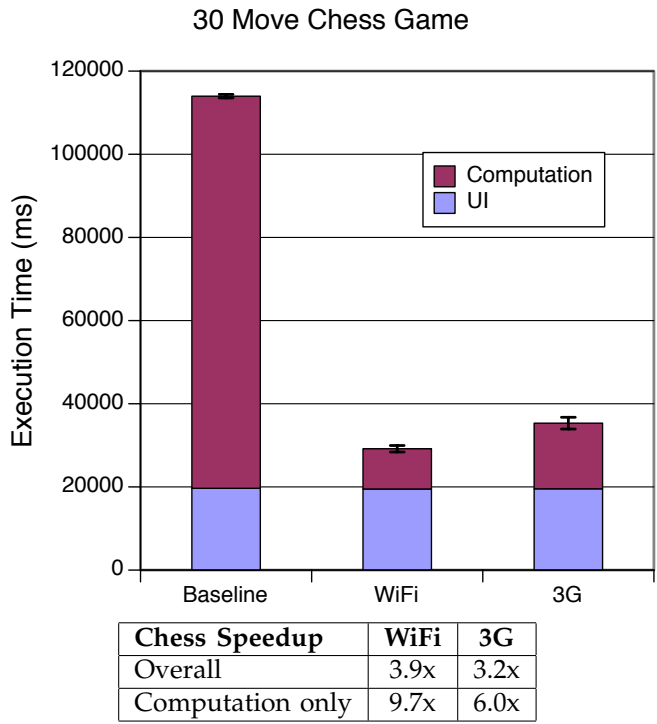}
\begin{tabular}{|l|c|c|} \hline
\textbf{Chess Speedup}&\textbf{WiFi}&\textbf{3G}\\ \hline
Overall     &3.9x &3.2x\\ \hline
Computation only  &9.7x &6.0x\\
\hline\end{tabular}
\vspace{2 mm}
\caption{\small\textsl{Absolute execution times for Pocket Chess with no offloading, WiFi offloading, and 3G offloading.  Each bar represents the average of 10 trials. Computation speed-up figures are shown in the table.}}
\label{fig:chessPerf}
\end{figure}

\begin{figure}
\centering
\epsfig{file=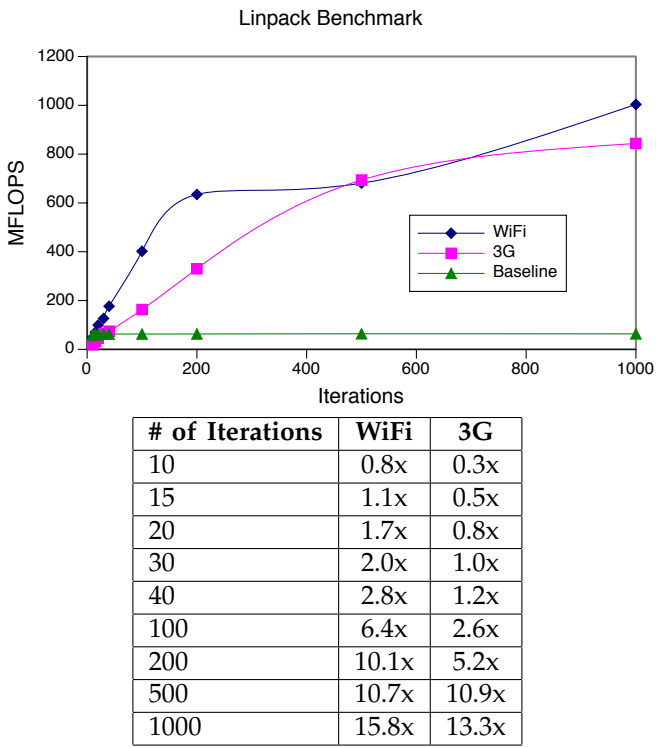, width=\columnwidth}
\begin{tabular}{|l|c|c|} \hline
\textbf{\# of Iterations}&\textbf{WiFi}&\textbf{3G}\\ \hline
10    &0.8x &0.3x\\ \hline
15    &1.1x &0.5x\\ \hline
20    &1.7x &0.8x\\ \hline
30    &2.0x &1.0x\\ \hline
40    &2.8x &1.2x\\ \hline
100   &6.4x &2.6x\\ \hline
200   &10.1x      &5.2x\\ \hline
500   &10.7x      &10.9x\\ \hline
1000  &15.8x      &13.3x\\
\hline\end{tabular}
\vspace{2 mm}
\caption{\small\textsl{MFLOPS for Linpack benchmark with no offloading, WiFi offloading, and 3G offloading by number of iterations of the benchmark. Each data point represents the average of 3 trials. Computation speed-up figures are shown in the table.}}
\label{fig:benchmark_mflops}
\end{figure}

\subsubsection{Linpack Benchmark}
We evaluated the performance of the Linpack benchmark, available as part of the 0xbench Google Play application \cite{0xbench_linpack:Online}.  The benchmark measures the system's floating point computing power by solving a dense \emph{n} x \emph{n} system of linear equations \cite{0xbench_linpack_wiki:Online}.

The benchmark runs through a computation-intensive loop 10 times. Without enabling COARA, the benchmark runs for 11.8ms on the Nexus 4, resulting in 58.7 MFLOPS. Because of network latency, enabling COARA results in slowdowns of 0.8x for WiFi and 0.3x for 3G.  To minimize the effect of network overhead, we ran the benchmark with a large number of iterations.  The results appear in Figure \ref{fig:benchmark_mflops}.  For 1000 iterations, we see speedups of 15.8x for WiFi and 13.3x 3G.  The benchmark's results demonstrate what our intuition tells us --- the usefulness of offloading is highly dependent on the size of the computation.

When evaluating the effect of COARA on battery consumption we observed a similar trend. At a low number of iterations, we find similar or worse battery consumption with offloading. As the number of iteration grows, COARA achieves energy consumption improvements of 34.5x over WiFi and 15.6x over 3G.  For lack of space, an energy consumption graph for Linpack was not included in this version of the paper.

\begin{figure*}
\centering
\epsfig{file=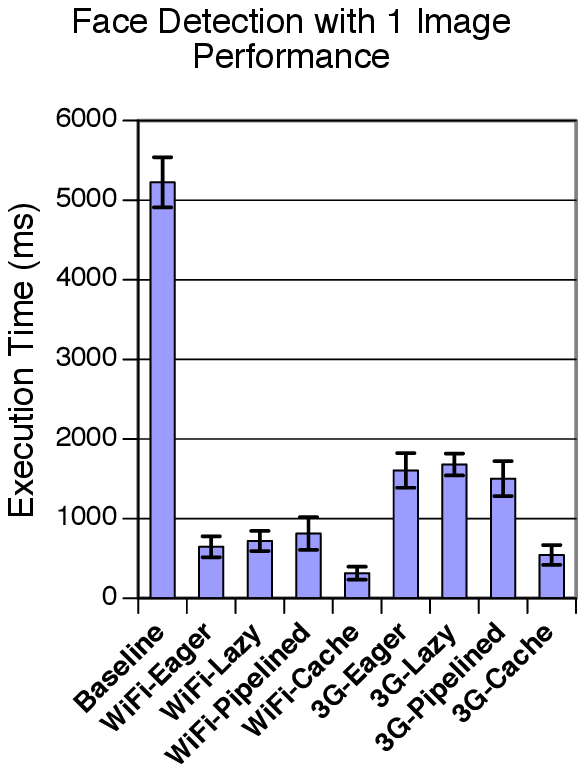,width=165pt}
\epsfig{file=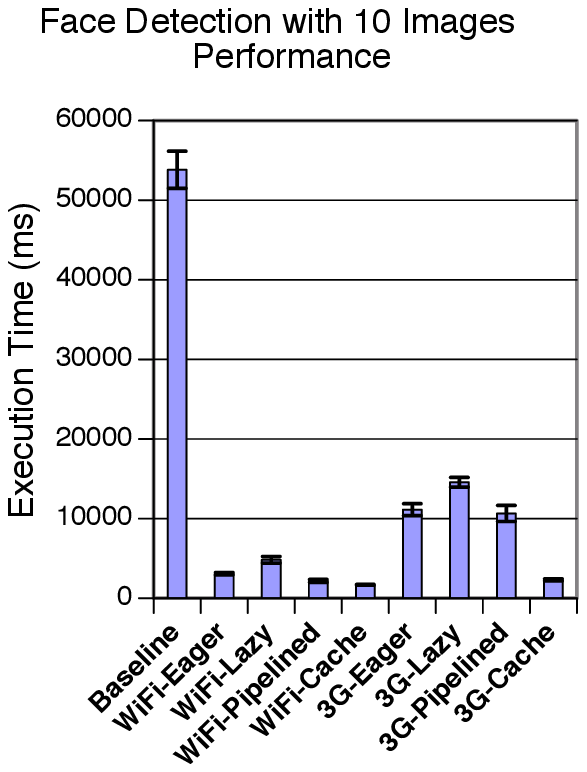,width=165pt}
\\
\begin{tabular}{|l|c|c|} \hline
\textbf{FD 1 Image}&\textbf{WiFi}&\textbf{3G}\\ \hline
Eager &8.1x &3.3x\\ \hline
Lazy  &7.3x &3.1x\\ \hline
Pipelined &6.4x &3.5x\\ \hline
Cache &16.6x  &9.6x\\
\hline\end{tabular}
\hspace{8 mm}
\begin{tabular}{|l|c|c|} \hline
\textbf{FD 10 Images}&\textbf{WiFi}&\textbf{3G}\\ \hline
Eager &17.5x  &4.8x\\ \hline
Lazy  &11.1x  &3.7x\\ \hline
Pipelined &24.7x  &5.1x\\ \hline
Cache &31.9x  &23.3x\\
\hline\end{tabular}
\vspace{2 mm}
\caption{\small\textsl{Absolute execution times for Face Detection with no offloading, WiFi offloading, and 3G offloading. Each bar represents the average of 10 trials.Computation speed-up figures are shown in the table.}}
\label{fig:fd_performance}
\end{figure*}

\begin{figure*}
\centering
\epsfig{file=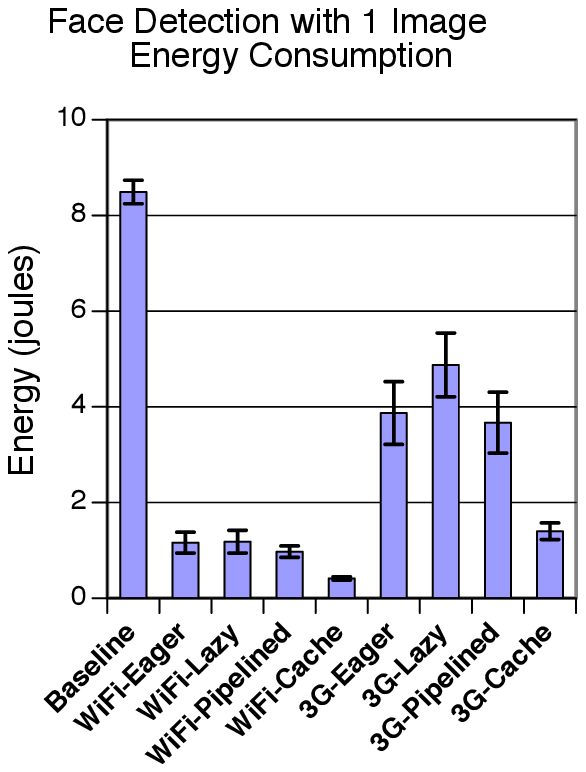,width=165pt}
\epsfig{file=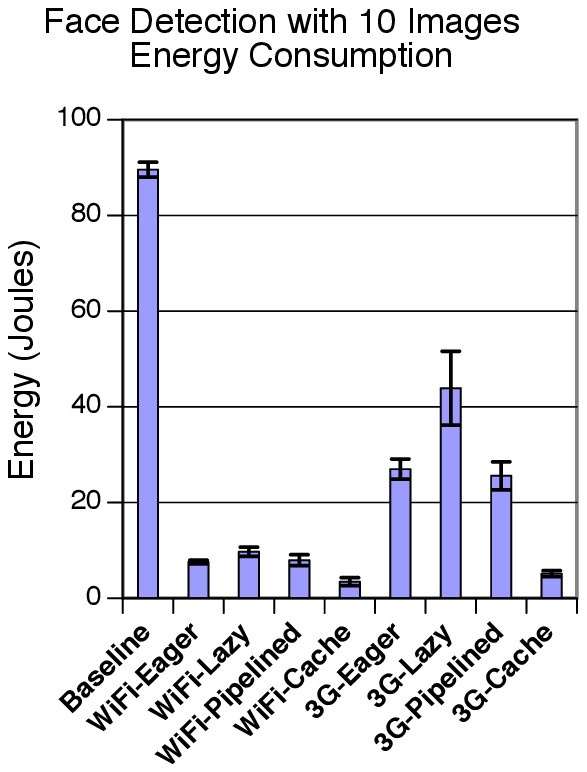,width=165pt}
\\
\begin{tabular}{|l|c|c|} \hline
\textbf{FD 1 Image Energy}&\textbf{WiFi}&\textbf{3G}\\ \hline
Eager &7.3x &2.2x\\ \hline
Lazy  &7.2x &1.7x\\ \hline
Pipelined   &8.7x &2.3x\\ \hline
Cache &20.5x      &6.1x\\
\hline\end{tabular}
\hspace{8 mm}
\begin{tabular}{|l|c|c|} \hline
\textbf{FD 10 Images Energy}&\textbf{WiFi}&\textbf{3G}\\ \hline
Eager &11.8x      &3.3x\\ \hline
Lazy  &9.2x &2.0x\\ \hline
Pipelined   &11.3x      &3.5x\\ \hline
Cache &25.9x      &17.4x\\
\hline\end{tabular}
\vspace{2 mm}
\caption{\small\textsl{Absolute energy consumption for Face Detection with no offloading, WiFi offloading, and 3G offloading. Each bar represents the average of 10 trials. Energy consumption improvement figures are shown in the table.}}
\label{fig:fd_energy}
\end{figure*}

\begin{figure}
\centering
\epsfig{file=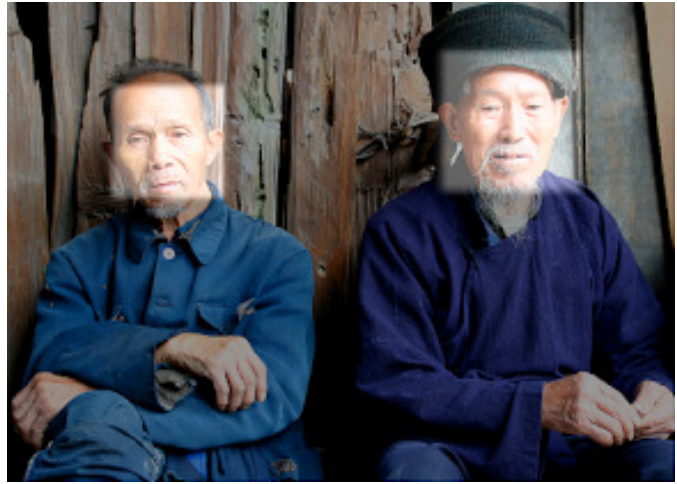, width=\columnwidth}
\caption{\small\textsl{Face detection with JJIL}}
\label{fig:two_faces}
\end{figure}

\begin{figure}
\centering
\epsfig{file=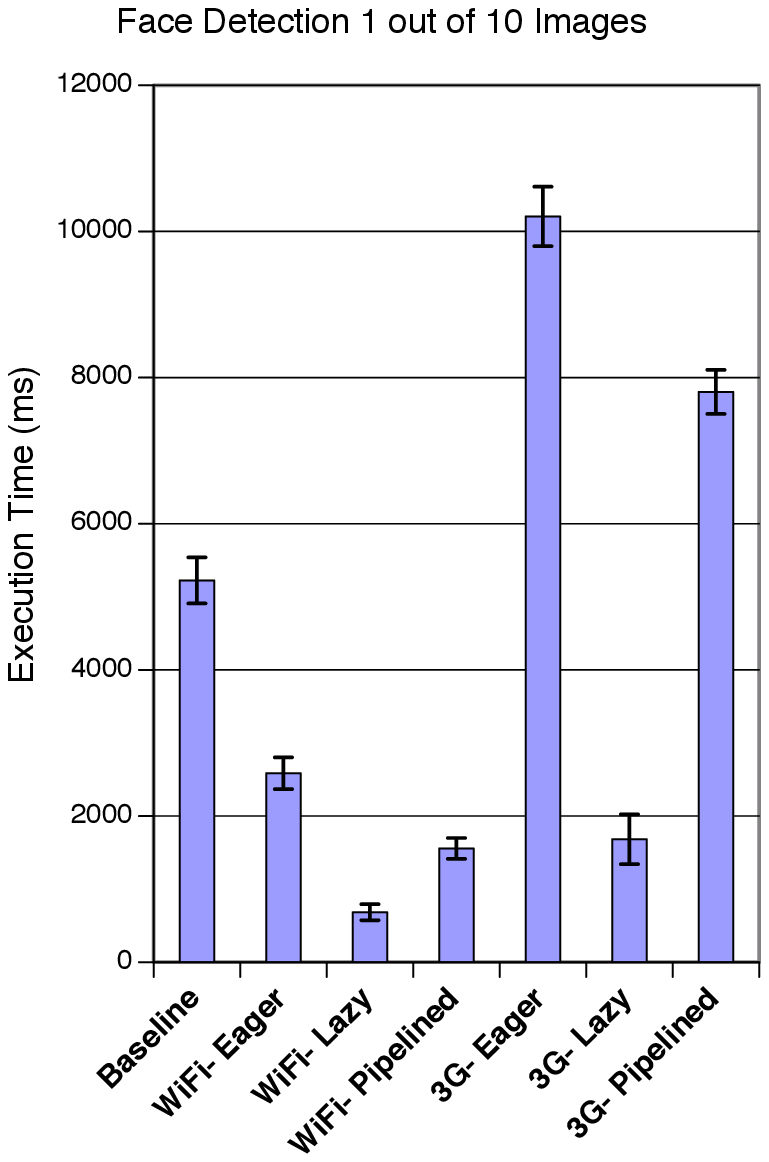, width=165pt}
\begin{tabular}{|l|c|c|} \hline
\textbf{FD 1/10 Images}&\textbf{WiFi}&\textbf{3G}\\ \hline
Eager &2.0x &0.5x\\ \hline
Lazy  &7.6x &3.1x\\ \hline
Pipelined &3.4x &0.7x\\
\hline\end{tabular}
\vspace{2 mm}
\caption{\small\textsl{Face Detection where the reachable heap contains 10 images yet only one is accessed by the offloaded method.  Absolute execution times with no offloading, WiFi offloading, and 3G offloading. Each bar represents the average of 10 trials. Computation speed-up figures are shown in the table.}}
\label{fig:fd_last_lazy}
\end{figure}

\subsubsection{Face Detection}

Thus far we have only looked at applications with negligible state transfer.  To demonstrate COARA's ability to optimize state transfer, we implemented a Face Detection Android application (Figure \ref{fig:two_faces}) that uses Jon's Java Image Library (JJIL) \cite{jjil_jon:Online}, an open source, pure Java image processing library.

We ran the Face Detection application with 1 and 10 images with \emph{eager}, \emph{pipelined}, and \emph{lazy transmission} via \emph{object proxies}.  When the cache is used, we preload the cache and conduct the tests with a best case 100\% hit rate.

\textbf{Face Detection with 1 Image}

The best performance for processing 1 image without the cache is achieved by \emph{eager transmission} which results in speedups of 8.1x for WiFi and 3.3x for 3G (see Figure \ref{fig:fd_performance}).  \emph{Eager transmission} does not use \emph{object proxies}.  Since only one large object is being transferred, there is no benefit to the overhead of object proxies.  We see the impact of this overhead with slightly lower speedups for \emph{lazy} and \emph{pipelined transmission}. We immediately see the benefits of the object cache with speedups of 16.6x for WiFi and 9.6x for 3G.

\textbf{Face Detection with 10 Images}

As COARA sends more large objects, \emph{eager transmission} yields greater performance.  Improvement occurs because the one time overhead of the initial offload is offset by the relatively larger gains from offloading.  \emph{Lazy transmission} does not see a significant improvement over 1 image because it now requires 10 calls from the server to client to retrieve each image, resulting in additional overhead.

The most dramatic improvement is seen by \emph{pipelined transmission}, which is able to take full advantage of the 10 images and provides a speedup of 24.7x over WiFi and 5.1x over 3G.  Another significant observation is that \emph{pipelined transmission} achieves a speedup of 1.42x WiFi relative to \emph{eager transmission}.  We feel this makes a powerful argument for the use of pipelined transmission.  However, pipelined transmission performed similarly to eager transmission over 3G.  Again, the object cache sees the best performance with speedups of 31.9x for WiFi and 23.3x for 3G.

\textbf{Face Detection and Energy Consumption}

The results in Figure \ref{fig:fd_energy} demonstrate that offloading with COARA significantly reduces energy consumption by up to 11.3x over WiFi and 3.5x over 3G without a cache, and 25.9x over WiFi and 17.4x over 3G with a cache.  However, we see no significant advantage with pipelined transmission over eager transmission in regards to energy consumption.

\textbf{Why use lazy transmission?}

In the previous examples, the offloaded method accessed all of the large objects in the reachable heap.  However, this may not be the case in practice.  In order to demonstrate this use case, we ran an experiment with 10 images in the reachable heap. However, in this experiment, the offloaded method performs face detection only on the 10th image in the array.  We will call this use case ``1/10'' face detection. The results are available in Figure \ref{fig:fd_last_lazy}.

With offloading over 3G, \emph{eager} and \emph{pipelined transmission} incur serious slowdowns of 0.5x and 0.7x respectively, eliminating the benefit from offloading.  The slowdowns occur because COARA transfers 10 large images, but code executing on the server only accesses one of them.

However, \emph{lazy transmission} achieves a speedup of 3.1x over 3G.  This is due to the fact that lazy transmission only retrieves the image from the client once it is accessed by code executing on the server.

In terms of energy consumption, eager and pipelined transmission both fail to
show any improvement over WiFi and 3G. However, lazy transmission improves energy consumption by 6.4x over WiFi and 1.8x over 3G. For lack of space, an energy consumption graph for 1/10 face detection was not included in this version of the paper.

We could have avoided this specific use case by modifying the application code to remove the unnecessary 9 images from the reachable state. However, sometimes it may be unclear which objects will be accessed by the offloaded method.  For example, conducting a search on a series of sorted objects.

\subsubsection{The Amount of State Transfer}

Table \ref{table:transmission} displays the amount of data transferred when running COARA.  It displays the amount of data uploaded from the device to the server and the amount of data downloaded from the server back to the mobile device.  We do not consider the transmission of the code at startup because in most cases the code will already be cached on the offloading server.

Benchmark and Chess have very low state transfer.  We did not enable object proxies with these two applications.  By having the application developer decide where to offload, we can reduce the number of unnecessary reachable objects in the heap leading to a smaller state transfer.

The table clearly illustrates the benefits of lazy transmission with the ``1/10'' face detection use case.  While both eager and pipelined transmission uploaded nearly 1.5MB to the server, lazy transmission only uploaded 117KB --- a reduction of over 90\%.  Similarly, the object cache kept uploads to the server below 2KB, thus reducing state transfer by up to 99\%.

\subsubsection{Alternative Code Execution}

As discussed in Section \ref{section.alternative_code_execution}, COARA provides the ability to increase fidelity by specifying alternative methods to run on the server.  To demonstrate this feature, we wrote a simple Android application that computes $\pi$ to a large number of digits with Machin's formula using code found in a Java RMI tutorial \cite{pi:Online}.

Our application runs a method \emph{pi()} that calculates $\pi$ to 10,000 digits.  Using COARA's alternative method feature, we specified an alternative method \emph{piServer()} that calculates $\pi$ to 20,000 digits.  COARA yielded a $\sim$5x speedup over WiFi and 3G and a higher quality result of twice as many digits.

\section{Limitations and Future Work}
\label{sec:limits}

While the focus of COARA is on improving performance, our top priority is to measure the impact of COARA on battery consumption.

COARA currently does not handle transferring of object locks.  This means that COARA assumes that only one method is offloaded at a time and that no other threads are running on the client in the meantime.  If this assumption is violated, object locks may not be respected which can result in unexpected behavior or deadlock.  COMET \cite{Gordon:2012vu} solves this problem with Distributed Shared Memory (DSM).

COARA is currently unable to offload native methods. \cite{Cuervo:2010:MMS:1814433.1814441} and \cite{Gordon:2012vu} have been able to accomplish this.  COARA could be extended to support this functionality.

COARA could be extended to use static analysis to detect which objects in the reachable heap are not accessed by a method and automatically annotate their classes with \emph{@EnableProxy}.  Static analysis could also be used to detect at compile time which objects are in the reachable heap and therefore must belong to classes that are Serializable.

Currently our decision engine is very simple.  We would like to develop an engine that determines not only when methods are offloaded, but also which offloading strategy to use.

Rather than define AspectJ annotations in the source code, it is possible to externalize the AspectJ to a configuration file, thereby leaving the source code untouched.  This would allow developers to use COARA with an ever smaller footprint.

We would like to explore offloading coarse-grained parallel algorithms where each machine operates on a copy of shared memory but is oblivious of changes on other machines.  An example would be performing facial recognition on multiple images in parallel.

\begin{table}
\centering
\caption{Total State Transfer\label{table:transmission}}
\begin{tabular}{|l|c|c|} \hline
\textbf{Application}&\textbf{Upload}&\textbf{Download}\\ \hline
Benchmark &6.4  &4.7\\ \hline
Chess (30 move game)  &83.5 &75.6\\ \hline
FD 1 Image Eager  &196.7  &51.5\\ \hline
FD 1 Image Lazy &196.7  &51.4\\ \hline
FD 1 Image Pipelined  &196.3  &50.7\\ \hline
FD 1 Image Cache  &1.3  &2.7\\ \hline
FD 10 Images Eager  &1486.9 &327.8\\ \hline
FD 10 Images Lazy &1493.6 &337.3\\ \hline
FD 10 Images Pipelined  &1495.6 &327.7\\ \hline
FD 10 Images Cache  &1.8  &7.8\\ \hline
FD 1/10 Eager &1487.5 &323.3\\ \hline
FD 1/10 Lazy  &117  &28.6\\ \hline
FD 1/10 Pipelined &1494.6 &62.1\\ \hline
\multicolumn{3}{|l|}{*FD = Face Detection} \\
\hline\end{tabular}
\end{table}

\section{Conclusion}
\label{sec:conclusion}

In this paper we have introduced COARA, a middleware platform for code offloading on Android that uses aspect-oriented programming (AOP) with AspectJ.  AOP allows COARA to intercept code for offloading without a custom compiler or modifying the operating system. COARA requires minimal changes to application source code and does not require the application developer to be aware of AspectJ.

Using AOP, COARA intercepts the transmission of large objects from the client and replaces them with proxies, thereby improving performance, reducing energy consumption, and minimizing state transfer.

We have shown that our pipelined transmission strategy was able to provide speedups of 24.7x over WiFi and 5.1x over 3G relative to execution on the mobile device.  In addition, pipelined transmission was able to achieve a 1.43x speedup over WiFi relative to eager transmission.

In one case study, we observed that using eager transmission with a large state transfer resulted in a slowdown of 0.5x over 3G.  The lazy transmission strategy minimized the state transfer by 90\% resulting in a speedup of 3x over 3G.  Lazy transmission can allow gains from offloading where they would be impossible with a full state transfer.

The object cache eliminates the need to resend unmodified objects.   When the cache hit rate is 100\%, COARA achieves speedups of 31.1x WiFi / 22.7x 3G and can decrease state transfer by 99\%.

COARA demonstrates that AOP offers an elegant solution for offloading. Using AOP, we implement techniques that mitigate the overhead of state transfer, which is one of the biggest impediments to the success of offloading.

\bibliographystyle{abbrv}
\bibliography{coara}

\begin{thebibliography}{10}

\bibitem{0xbench_linpack:Online}
0xbench android benchmarking application. https://play.google.com/store/apps/
  details?id=org.zeroxlab.zeroxbenchmark.

\bibitem{SSHTunnel:Online}
Android ssh tunneling. https://play.google.com/store/apps/
  details?id=org.sshtunnel.

\bibitem{androidx86:Online}
Android-x86 - porting android to x86. http://www.android-x86.org/.

\bibitem{aspectj:Online}
Aspectj aspect-oriented java extension. http://eclipse.org/aspectj/.

\bibitem{gson:Online}
Google-gson - a java library to convert json to java object.
  https://code.google.com/p/google-gson/.

\bibitem{pi:Online}
Java rmi tutorial. http://docs.oracle.com/javase/tutorial/rmi/
  examples/client/pi.java.

\bibitem{jjil_jon:Online}
Jjil - jon's java imaging library. https://code.google.com/p/jjil/.

\bibitem{kryo:Online}
Kryo - fast, efficient java serialization and cloning.
  https://code.google.com/p/kryo/.

\bibitem{0xbench_linpack_wiki:Online}
Linpack benchmarks - wikipedia.
  https://en.wikipedia.org/wiki/linpack\_benchmarks.

\bibitem{lipermi:Online}
Lipermi rmi implementation. http://lipermi.sourceforge.net/.

\bibitem{pocketChess:Online}
Pocket chess for android. https://play.google.com/store/apps/
  details?id=kobi.chess.

\bibitem{uiautomator:Online}
Uiautomator testing framework. http://developer.android.com/tools/
  help/uiautomator.

\bibitem{xstream:Online}
Xstream serialization library. http://xstream.codehaus.org/.

\bibitem{Balan:2003uz}
R.~K. Balan, M.~Satyanarayanan, S.~Y. Park, and T.~Okoshi.
\newblock {Tactics-based remote execution for mobile computing}.
\newblock In {\em MobiSys '03: Proceedings of the 1st international conference
  on Mobile systems, applications and services}. ACM Request Permissions, May
  2003.

\bibitem{Charles:2005ud}
P.~Charles, C.~Grothoff, V.~Saraswat, C.~Donawa, A.~Kielstra, K.~Ebcioglu,
  C.~Von~Praun, and V.~Sarkar.
\newblock {X10: an object-oriented approach to non-uniform cluster computing}.
\newblock {\em ACM Sigplan Notices}, 40(10):519--538, 2005.

\bibitem{Chen_Androidx86:2010bx}
E.~Y. Chen, M.~W. o. W.~M. Itoh, and M.~N. W. . I. I.~S. on~a.
\newblock {Virtual smartphone over IP}.
\newblock {\em World of Wireless Mobile and Multimedia Networks (WoWMoM), 2010
  IEEE International Symposium on a}, 2010.

\bibitem{Chen:2012cs}
H.-Y. Chen, Y.-H. Lin, and C.-M. Cheng.
\newblock {COCA: Computation Offload to Clouds using AOP}.
\newblock In {\em 2012 12th IEEE/ACM International Symposium on Cluster, Cloud
  and Grid Computing (CCGrid)}, pages 466--473. IEEE, Sept. 2012.

\bibitem{Chun:2011ty}
B.~G. Chun, S.~Ihm, P.~Maniatis, M.~Naik, and A.~Patti.
\newblock {Clonecloud: elastic execution between mobile device and cloud}.
\newblock {\em Proceedings of the sixth conference on Computer systems}, pages
  301--314, 2011.

\bibitem{Cuervo:2010:MMS:1814433.1814441}
E.~Cuervo, A.~Balasubramanian, D.-k. Cho, A.~Wolman, S.~Saroiu, R.~Chandra, and
  P.~Bahl.
\newblock {MAUI: making smartphones last longer with code offload}.
\newblock In {\em MobiSys '10: Proceedings of the 8th international conference
  on Mobile systems, applications, and services}, pages 49--62, New York, NY,
  USA, June 2010. ACM Request Permissions.

\bibitem{Fahim:2013dd}
A.~Fahim, A.~Mtibaa, and K.~A. Harras.
\newblock {Making the case for computational offloading in mobile device
  clouds}.
\newblock In {\em MobiCom '13: Proceedings of the 19th annual international
  conference on Mobile computing {\&} networking}. ACM, Sept. 2013.

\bibitem{Flinn:2012ws}
J.~Flinn.
\newblock Cyber foraging: Bridging mobile and cloud computing.
\newblock {\em Synthesis Lectures on Mobile and Pervasive Computing},
  7(2):1--103, 2012.

\bibitem{Flinn:2001wh}
J.~Flinn, D.~Narayanan, and M.~Satyanarayanan.
\newblock {Self-tuned remote execution for pervasive computing}.
\newblock pages 61--66, 2001.

\bibitem{DesignPatterns:1995wx}
E.~Gamma, R.~Helm, R.~Johnson, and J.~Vlissides.
\newblock {\em {Design patterns: elements of reusable object-oriented
  software}}.
\newblock Addison-Wesley Longman Publishing Co., Inc, 1995.

\bibitem{Giurgiu:2009vf}
I.~Giurgiu, O.~Riva, D.~Juric, I.~Krivulev, and G.~Alonso.
\newblock {Calling the cloud: Enabling mobile phones as interfaces to cloud
  applications}.
\newblock {\em Middleware 2009}, pages 83--102, 2009.

\bibitem{Gordon:2012vu}
M.~S. Gordon, D.~A. Jamshidi, S.~Mahlke, Z.~M. Mao, and X.~Chen.
\newblock {COMET: code offload by migrating execution transparently}.
\newblock In {\em OSDI'12: Proceedings of the 10th USENIX conference on
  Operating Systems Design and Implementation}. USENIX Association, Oct. 2012.

\bibitem{Ha_JustInTime:2012wx}
K.~Ha, P.~Pillai, W.~Richter, Y.~Abe, and M.~Satyanarayanan.
\newblock {Just-in-Time Provisioning for Cyber Foraging}.
\newblock {\em CMU School of Computer Science, Tech. Rep. CMU-CS-12-148}, 2012.

\bibitem{Hericko:2003tk}
M.~Hericko, M.~B. Juric, I.~Rozman, S.~Beloglavec, and A.~Zivkovic.
\newblock {Object serialization analysis and comparison in java and. net}.
\newblock {\em ACM Sigplan Notices}, 38(8):44--54, 2003.

\bibitem{Holder:1999wj}
O.~Holder, I.~Ben-Shaul, and H.~Gazit.
\newblock {Dynamic layout of distributed applications in FarGo}.
\newblock {\em Proceedings of the 21st international conference on Software
  engineering}, pages 163--173, 1999.

\bibitem{Kemp:2012vn}
R.~Kemp, N.~Palmer, T.~Kielmann, and H.~Bal.
\newblock {Cuckoo: a computation offloading framework for smartphones}.
\newblock {\em Mobile Computing, Applications, and Services}, pages 59--79,
  2012.

\bibitem{AOP_Paper:1997vp}
G.~Kiczales, J.~Lamping, A.~Mendhekar, C.~Maeda, C.~Lopes, J.-M. Loingtier, and
  J.~Irwin.
\newblock {Aspect-oriented programming}.
\newblock 1997.

\bibitem{Kosta:2012vx}
S.~Kosta, A.~Aucinas, P.~Hui, R.~Mortier, and X.~Zhang.
\newblock {Thinkair: Dynamic resource allocation and parallel execution in the
  cloud for mobile code offloading}.
\newblock {\em INFOCOM, 2012 Proceedings IEEE}, pages 945--953, 2012.

\end{thebibliography}


%
%
%

\end{document}